\documentclass[12pt]{article}
\usepackage{psfrag}
\usepackage{graphicx}
\usepackage{graphics}
\newcommand{\as}{a\!\!\!/}
\newcommand{\As}{A\!\!\!/}
\newcommand{\ks}{k\!\!\!/}

\date{}
\begin{document}
\baselineskip=18.6pt plus 0.2pt minus 0.1pt \makeatletter


\title{\vspace{-3cm}
\hfill\parbox{4cm}{\normalsize \emph{LPHEA 04-08}}\\
 \vspace{1cm}
{Relativistic Electronic Dressing in Laser-assisted ionization of
atomic hydrogen by electron impact.}}
 \vspace{2cm}

\author{Y.  Attaourti$^{\dag}$\thanks{attaourti@ucam.ac.ma}, S. Taj$^\ddag$\\
 {\it {\small $^\dag$Laboratoire de Physique des Hautes
Energies et d'Astrophysique, Facult\'e }}\\ {\it {\small des
Sciences Semlalia, Universit\'e Cadi Ayyad Marrakech, BP : 2390,
Maroc.}}
\\
{\it {\small $^\ddag$UFR de Physique Atomique Mol\'eculaire et
Optique Appliqu\'ee,
 Facult\'e des Sciences,}}\\
{\it {\small Universit\'e Moulay Isma\"{\i}l
   BP : 4010, Beni M'hamed, Mekn\`es, Maroc.}}}\maketitle \setcounter{page}{1}
\begin{abstract}
Within the framework of the coplanar binary geometry where it is
justified to use plane wave solutions for the study of the
$(e,2e)$ reaction and in the presence of a circularly polarized
laser field, we introduce as a first step the DVRPWBA1
(Dirac-Volkov Plane Wave Born Approximation1) where we take into
account only the relativistic dressing of the incident and
scattered electrons. Then, we introduce the DVRPWBA2 (Dirac-Volkov
Plane Wave Born Approximation2) where we take totally into account
the relativistic dressing of the incident, scattered and ejected
electrons. We then compare the corresponding triple differential
cross sections for laser-assisted ionization of atomic hydrogen by
electron impact both for the non relativistic and the relativistic
regime.\\ \vspace{.04cm}\\
 PACS number(s): 34.80.Qb, 12.20.Ds
\end{abstract}

\maketitle
\section{Introduction}
Ehrhardt \textit{et al} (1969) \cite{1} were the first to conduct
electron impact ionization experiments in which the two outgoing
electrons are detected in coincidence after angular and energy
analysis. The first theoreticians who proposed such type of
experiment were Smirnov and Neudachin (1966) \cite{2}.
These experiments now are commonly referred to as $(e,2e)$. Since then, the $%
(e,2e)$ reaction has been studied extensively in the non
relativistic kinematic domain (Camilloni \textit{et al} 1972 \cite{3},
Weigold \textit{et al} 1973 \cite{4}, van der Wiel 1973 \cite{5}, Brion
1975 \cite{6}). All these studies show that the non relativistic
(e,2e) reaction is a very sensitive test of the target electronic
structure and the electron impact ionization reaction mechanism.
For the relativistic domain, the first electron impact ionization
experiments where conducted by Dangerfield and Spicer (1975) \cite{7},
then by Hoffman \textit{et al} (1979) \cite{8} and by Anholt (1979)
\cite{9}. Total cross sections with relativistic electrons where
measured for the K and L shells of heavy elements. Some
theoreticians ( Scoffield 1978 \cite{10}, Moiseiwitsch and Stockman
\cite{11}) proposed models for total ionization cross sections.
Finally, Fuss \textit{et al} (1982) \cite{12} proposed a model of an
$(e,2e)$ reaction they called binary $(e,2e)$ reaction in which
the maximum momentum transfer occurs, that is, a reaction where
the outgoing electrons have equal energy. This type of reaction
has been since then the most successful for probing atomic,
molecular and solid state structure. In a report devoted both to
experimental and theoretical developments in the study of
relativistic $(e,2e)$ processes, Nakel and Whelan (1999) \cite{13}
reviewed the goals of these investigations aimed at gaining a
better understanding of inner-shell ionization process by
relativistic electrons up to the highest atomic numbers and
probing the quantum mechanical Coulomb problem in the regime of
high energies (up to 500 keV) and strong fields. With the advent
of the laser field, many theoretical models have been proposed
\cite{14} mainly in the non relativistic domain whereas in the
relativistic domain we can only quote the work of Reiss (1990)
\cite{15} and that of Crawford and Reiss (1994,1998) \cite{16} who studied
the relativistic ionization of hydrogen (without electron impact)
by linearly polarized light. They focused their work on the
calculations of the differential transition rates and have shown
that strong field atomic stabilization is enhanced by
relativistic effects.

In this article, we present a theoretical model for the
relativistic electronic dressing in laser-assisted ionization of
atomic hydrogen by electron impact with a circularly polarized
laser field and in order to check the consistency of our
calculations, we begin our study in absence of the laser field.
As we devote this analysis to atomic hydrogen, all distorsion
effects mentioned in \cite{13} need not be addressed since the atomic
number we deal with is that of atomic hydrogen, that is $Z=1$. It
is checked first that , in the absence of the laser field and
working in the coplanar binary geometry where the kinetic
energies of the scattered electron and the ejected electron are
nearly the same, it is justified to use plane wave solutions for
the study of the $(e,2e)$ reaction. Indeed, this particular
geometry is such that the ejected electron does not feel the
Coulomb influence of atomic target and can be described by a
plane wave in the non relativistic as well as in the relativistic
domain. Then, in the presence of a circularly polarized laser
field, we introduce as a first step the DVRPWBA1 (Dirac-Volkov
Plane Wave Born Approximation1) where we take into account only
the relativistic dressing of the incident and scattered
electrons. It is shown that this approximation introduces an
asymmetry in the description of the scattering process since it
does not allow photon exchange between the laser field and the
ejected electron and its domain of validity is only restricted to
very weak fields and non relativistic electron kinetic energy.
Then, we introduce the DVRPWBA2 (Dirac-Volkov Plane Wave Born
Approximation2) where we take totally into account the
relativistic dressing of the incident, scattered and ejected
electrons.

The organization of this paper is as follows : in section II, we
present the relativistic formalism of $(e,2e)$ reaction in
absence of the laser field RPWBA (Relativistic Plane Wave Born
Approximation) and we compare it in the non relativistic domain
with the NRPWBA (Non Relativistic Plane wave Born Approximation)
as well as the NRCBA (Non Relativistic Coulomb-Born
Approximation). In section III, we introduce the DVPWBA1
(Dirac-Volkov Plane Wave Born Approximation1) . This
approximation is introduced as a first step. In section IV, we
introduce the DVPWBA2 (Dirac-Volkov Plane Wave Born
Approximation2) in which we take fully account of the
relativistic electronic dressing of the incident, scattered and
ejected electrons. This approximation is more founded on physical
grounds since the ejected electron can also exchange photons
(absorption or emission) with the laser field. This more complete
description of the incoming and outgoing electrons allows to
investigate the relativistic domain. In section V, we discuss the
results we have obtained and we end by a brief conclusion in
Section VI. Throughout this work, atomic units (a.u) are used
($\hbar =m_{e}=e=1$) where $m_{e}$ is the electron mass and TDCS
stands for triple differential cross section.

\section{The TDCS in absence of the laser field}

The transition matrix element for the direct channel (we neglect
exchange effects) is given by

\begin{eqnarray}
&&S_{fi}=-\frac{i}{c}\int_{-\infty }^{\infty }dx^{0} <\psi
_{p_{f}}(x_{1})\phi _{f}(x_{2})\mid V_{d}\mid \psi
_{p_{i}}(x_{1})\phi _{i}(x_{2})>,
\end{eqnarray}
where $V_{d}=1/r_{12}-1/r_{1}$ is the direct interaction potential ($%
t_{1}=t_{2}=t\Longrightarrow x_{1}^{0}=x_{2}^{0}=x^{0}$) and in the RPWBA, $%
\psi _{p_{f}}(x_{1})$ is the wave function describing the
scattered electron

\begin{equation}
\psi
_{p_{f}}(x_{1})=\frac{u(p_{f},s_{f})}{\sqrt{2E_{f}V}}e^{-ip_{f}.x_{1}},
\end{equation}
given by a free Dirac solution normalized to the volume $V$. For
the incident electron, we use

\begin{equation}
\psi
_{p_{i}}(x_{1})=\frac{u(p_{i},s_{i})}{\sqrt{2E_{i}V}}e^{-ip_{i}.x_{1}}.
\end{equation}
For the atomic target, $\phi _{i}(x_{2})=\phi
_{i}(t,\mathbf{r}_{2})$ is the relativistic wave function of
atomic hydrogen in its ground state. For the ejected electron, we
use again a free Dirac solution normalized to the volume $V$ and
$\phi _{f}(x_{2})$ is given by

\begin{equation}
\phi _{f}(x_{2})=\psi _{p_{B}}(x_{2})=\frac{u(p_{B},s_{B})}{\sqrt{2E_{B}V}}%
e^{-ip_{B}.x_{2}}.
\end{equation}
The free spinor $u(p,s)$ is such that $\overline{u}(p,s)u(p,s)=2c^{2}$ and $%
u^{\dagger }(p,s)u(p,s)=2E$. Using the standard methods of QED
[17], we obtain for the unpolarized TDCS

\begin{eqnarray}
\frac{d\overline{\sigma }}{dE_{B}d\Omega _{B}d\Omega _{f}} &=&\frac{%
|\mathbf{p}_{f}||\mathbf{p}_{B}|}{|\mathbf{p}_{i}|c^{2}}(\sum_{s_{B}}\overline{u}(p_{B},s_{B})u(p_{B},s_{B}))%
\frac{(2E_{i}E_{f}/c^{2}-p_{i}.p_{f}+c^{2})}{\mid \mathbf{p}_{f}\mathbf{-p}%
_{i}\mid ^{4}}\nonumber\\ &\times&  \mid\Phi
_{1,1/2,1/2}(\mathbf{q}_{1}\mathbf{=\Delta -p}_{B})-\Phi
_{1,1/2,1/2}(\mathbf{q}_{0}\mathbf{=-p}_{B})\mid ^{2}.
\end{eqnarray}

The sum over the spins of the ejected electron gives

\begin{equation}
\sum_{s_{B}}\overline{u}(p_{B},s_{B})u(p_{B},s_{B})=4E_{B}.
\end{equation}
The functions $\Phi _{1,1/2,1/2}(\mathbf{q})$ are the Fourier
transforms of the relativistic atomic hydrogen wave functions

\begin{eqnarray}
&&\Phi _{1,1/2,1/2}(\mathbf{q})=(2\pi )^{(-3/2)}
\int d\mathbf{r}_{2}e^{i%
\mathbf{q.r}_{_{2}}}\Psi _{n=1,j=1/2,m=1/2}(\mathbf{r}_{2}),
\end{eqnarray}
and $\mathbf{\Delta =p}_{i}-\mathbf{p}_{f}\ $is the momentum
transfer. This TDCS is to be compared with the corresponding one
in the NRPWBA (Non Relativistic Plane Wave Born Approximation)

\begin{eqnarray}
&&\frac{d\overline{\sigma }}{dE_{B}d\Omega _{B}d\Omega
_{f}}=\frac{2^{7}}{ (2\pi
)^{2}}\frac{|\mathbf{p}_{f}||\mathbf{p}_{B}|}{|\mathbf{p}_{i}|}\frac{1}{\mid
\mathbf{\Delta }\mid ^{4}}
\left\{\frac{1}{(\mathbf{q}_{1}^{2}+1)^{2}}-\frac{1}{(\mathbf{q}_{0}^{2}+1)^{2}}
\right\}^2,
\end{eqnarray}

where $\mathbf{q}_{1}=\mathbf{\Delta -p}_{B}$ and $\mathbf{q}_{0}=\mathbf{-p}%
_{B}$ to the TDCS in the NRCBA (Non Relativistic Coulomb Born
Approximation)

\begin{equation}
\frac{d\sigma ^{CB}}{dE_{B}d\Omega _{B}d\Omega _{f}}=\frac{|\mathbf{p}_{f}||\mathbf{p}_{B}|}{|\mathbf{p}_{i}|}%
\mid f_{ion}^{CB1}\mid ^{2},
\end{equation}
where $f_{ion}^{CB1}$ is the first Coulomb-Born amplitude
corresponding to the ionization of atomic hydrogen by electron
impact \cite{18}

\begin{equation}
f_{ion}^{CB1}=-\frac{2}{\mid \mathbf{\Delta }\mid
^{2}}M_{1s}(\mathbf{\Delta },\mathbf{p}_{B}).
\end{equation}
The quantity $M_{1s}(\mathbf{\Delta },\mathbf{p}_{B})$ is easily
deduced from the Nordsieck integral \cite{19}

\begin{eqnarray*}
I(\lambda ) &=&\int e^{-\lambda
r}\frac{e^{i\mathbf{q.r}}}{r}\quad1F1(
\frac{i}{p_{B}},1,i(p_{B}r+\mathbf{p}_{B}.\mathbf{r}))d\mathbf{r}
\\
&=&\frac{4\pi }{q^{2}+\lambda ^{2}}[\frac{q2+\lambda 2+2\mathbf{q.p}%
_{B}-2i\lambda p_{B}}{q^{2}+\lambda ^{2}}]^{-i/p_{B}},
\end{eqnarray*}
giving the well known result

\begin{equation}
M_{1s}(\mathbf{\Delta },\mathbf{p}_{B})=\frac{e^{\pi
/2p_{B}}}{2\sqrt{2}\pi ^{2}}\Gamma
(1-\frac{i}{p_{B}})(-\frac{dI(\lambda )}{d\lambda })_{\lambda =1}.
\end{equation}

\section{The TDCS in presence of the laser field. The DVPWBA1}

We begin our study of the (e,2e) reaction by considering first
the DVPWBA1 where we only take into account the dressing of the
incident and the scattered electron. The laser field is
circularly polarized. Again, the transition matrix element for
the direct channel is

\begin{eqnarray}
&&S_{fi}=-\frac{i}{c}\int_{-\infty }^{\infty }dx^{0}<\psi
_{q_{f}}(x_{1})\phi _{f}(x_{2})\mid V_{d}\mid \psi
_{q_{i}}(x_{1})\phi _{i}(x_{2})>,
\end{eqnarray}
where ($t_{1}=t_{2}=t\Longrightarrow x_{1}^{0}=x_{2}^{0}=x^{0}$)
and in the DVPWBA1, $\psi _{q_{f}}(x_{1})$ is the Dirac-Volkov
wave function normalized to the volume $V$ describing the
scattered electron

\begin{equation}
\psi _{q_{f}}(x_{1})=[1+\frac{\ks\As_{(1)}}{2c(k.p_{f})}]\frac{u(p_{f},s_{f})}{\sqrt{2Q_{f}V}}e^{is_{f}(x_{1})},
\end{equation}
where $A_{(1)}=a_{1}\cos (\phi _{1})+a_{2}\sin (\phi _{1})$ is
the four
potential of the laser field, $\phi _{1}=k.x_{1}=k_{0}x_{1}^{0}-\mathbf{k.x}%
_{1}=wt-\mathbf{k.x}_{1}$ is the phase of the laser field and $w$
its pulsation. $Q$ is the total energy acquired by the electron
in the presence of a laser field and is given by :

\begin{equation}
Q=E-\frac{a^{2}w}{2c^{2}(k.p)}.
\end{equation}
The phase $s_{f}(x_{1})$ is given by

\begin{equation}
s_{f}(x_{1})=-q_{f}.x_{1}-\frac{a_{1}.p_{f}}{c(k.p_{f})}\sin (\phi _{1})+%
\frac{a_{2}.p_{f}}{c(k.p_{f})}\cos (\phi _{1}).
\end{equation}
For the incident electron, we use

\begin{equation}
\psi _{q_{i}}(x_{1})=[1+\frac{\ks\As_{(1)}}{2c(k.p_{i})}]\frac{u(p_{i},s_{i})}{\sqrt{%
2Q_{i}V}}e^{is_{i}(x_{1})},
\end{equation}
with the phase $s_{i}(x_{1})$ given by

\begin{equation}
s_{i}(x_{1})=-q_{i}.x_{1}-\frac{a_{1}.p_{i}}{c(k.p_{i})}\sin (\phi _{1})+%
\frac{a_{2}.p_{i}}{c(k.p_{i})}\cos (\phi _{1}),
\end{equation}
where the four vector $q^{\mu }$ is such

\begin{equation}
q^{\mu }=p^{\mu }-\frac{a^{2}}{2c^{2}(k.p)}k^{\mu },
\end{equation}
and $a^{2}=A_{\mu }A^{\mu }=a_{1}^{2}=a_{2}^{2}$. For the atomic target, $%
\phi _{i}(x_{2})=\phi _{i}(t,\mathbf{r}_{2})=e^{-i\varepsilon
_{b}t}\phi _{i}(\mathbf{r}_{2})$ is the relativistic wave
function of atomic hydrogen in its ground state and $\varepsilon
_{b}=c^{2}(\sqrt{1-\alpha ^{2}}-1)$ is the binding energy of the
ground state of atomic hydrogen with $\alpha =1/c$ the fine
structure constant. For the ejected electron, we use a free Dirac
solution normalized to the volume $V$ and $\phi _{f}(x_{2})$

\begin{equation}
\phi _{f}(x_{2})=\psi _{p_{B}}(x_{2})=\frac{u(p_{B},s_{B})}{\sqrt{2E_{B}V}}%
e^{-ip_{B}.x_{2}}.
\end{equation}
Using the standard methods of QED, we have for the unpolarized
TDCS

\begin{equation}
\frac{d\overline{\sigma }}{dE_{B}d\Omega _{B}d\Omega
_{f}}=\left.\sum_{s=-\infty
}^{\infty }\frac{d\overline{\sigma }^{(s)}}{dE_{B}d\Omega _{B}d\Omega _{f}}%
\right|_{Q_{f}=Q_{i}+sw+\varepsilon _{b}-E_{B}},
\end{equation}
where the expression of $d\overline{\sigma }^{(s)}/dE_{B}d\Omega
_{B}d\Omega _{f}$ is

\begin{eqnarray}
\frac{d\overline{\sigma }^{(s)}}{dE_{B}d\Omega _{B}d\Omega _{f}}&=&
\frac{1}{%
2}\frac{|\mathbf{q}_{f}||\mathbf{p}_{B}|}{|\mathbf{q}_{i}|c^{6}}\frac{(\sum_{s_{i},s_{f}}\mid
M_{fi}^{(s)}\mid ^{2}/2)}{\mid \mathbf{q}_{f}\mathbf{-q}_{i}\mathbf{-}s%
\mathbf{k}\mid ^{4}}(4E_{B}) \nonumber\\ &\times& \mid \Phi
_{1,1/2,1/2}(\mathbf{q=\Delta }_{s}\mathbf{-p}_{B})-\Phi
_{1,1/2,1/2}(\mathbf{q=-p}_{B})\mid ^{2}.
\end{eqnarray}

The sum $(\sum_{s_{i},s_{f}}\mid M_{fi}^{(s)}\mid
^{2}/2)$ has already been
evaluated in a previous work \cite{20} and $\mathbf{\Delta }_{s}=\mathbf{q}_{i}%
\mathbf{-q}_{f}+s\mathbf{k}$ is the momentum transfer in presence
of the laser field. This TDCS is compared to the corresponding
TDCSs in the non relativistic regime. On the one hand, the
calculations within the framework of the NRPWBA1 (where the
incident and scattered electrons are described by non
relativistic Volkov plane waves whereas the ejected electron is
described by a non relativistic free plane wave) give

\begin{equation}
\frac{d\overline{\sigma }^{NRPWBA1}}{dE_{B}d\Omega _{B}d\Omega _{f}}%
\left.\sum_{s=-\infty }^{\infty }\frac{d\overline{\sigma
}^{(s)}}{dE_{B}d\Omega _{B}d\Omega _{f}}\right|_
{E_{f}=E_{i}+sw+\varepsilon _{1s}-E_{B}},
\end{equation}
with

\begin{eqnarray}
\frac{d\overline{\sigma }^{(s)}}{dE_{B}d\Omega _{B}d\Omega
_{f}}&=&\frac{2^{7} }{(2\pi
)^{2}}\frac{|\mathbf{p}_{f}||\mathbf{p}_{B}|}{|\mathbf{p}_{i}|}\frac{J_{s}^{2}(z_{NR})}{\mid
\mathbf{p }_{f}\mathbf{-p}_{i}\mathbf{-}s\mathbf{k}\mid ^{4}}
\left\{\frac{1}{(\mathbf{q}
_{1s}^{2}+1)^{2}}-\frac{1}{(\mathbf{q}_{0s}^{2}+1)^{2}}\right\}^2,
\end{eqnarray}
where $\varepsilon _{1s}=-0.5$ a.u is the non relativistic
binding energy of
atomic hydrogen in its ground state, $\mathbf{q}_{1s}=\mathbf{p}_{i}\mathbf{+%
}s\mathbf{k-p}_{f}-\mathbf{p}_{B}=\mathbf{\Delta
}_{s}^{NR}\mathbf{-p}_{B}$ and $\mathbf{q}_{0s}=-\mathbf{p}_{B}$.
On the other hand, the calculations within the framework of the
NRCBA1 (where the incident and scattered electrons are described
by non relativistic Volkov plane waves whereas the ejected
electron is described by a Coulomb wave function) give

\begin{equation}
\frac{d\overline{\sigma }^{NRCBA1}}{dE_{B}d\Omega _{B}d\Omega
_{f}} =\left.\sum_{s=-\infty }^{\infty }\frac{d\overline{\sigma
}^{(s)}}{dE_{B}d\Omega _{B}d\Omega _{f}}\right|_{
E_{f}=E_{i}+sw+\varepsilon _{1s}-E_{B}},
\end{equation}
with

\begin{eqnarray}
&&\frac{d\overline{\sigma }^{(s)}}{dE_{B}d\Omega _{B}d\Omega
_{f}}=\frac{1 }{2\pi
^{4}}\frac{|\mathbf{p}_{f}||\mathbf{p}_{B}|}{|\mathbf{p}_{i}|}\frac{J_{s}^{2}(z_{NR})}{\mid
\mathbf{p} _{f}\mathbf{-p}_{i}\mathbf{-}s\mathbf{k}\mid
^{4}}e^{\pi /p_{B}}\mid \Gamma (1\mathbf{-}\frac{i}{p_{B}})\mid
^{2}\mid I(\mathbf{q}_{s}=\mathbf{p}_{i}
\mathbf{-p}_{f}+s\mathbf{k-p}_{B})\mid ^{2}.
 \end{eqnarray}

Note that we may write $\mathbf{q}_{s}=\mathbf{p}_{i}\mathbf{-p}_{f}+s%
\mathbf{k-p}_{B}=\mathbf{\Delta }_{s}^{NR}\mathbf{-p}_{B}$. The
result for $ I(\mathbf{q}_{s}=\mathbf{\Delta
}_{s}^{NR}\mathbf{-p}_{B})$ is

\begin{eqnarray}
I(\mathbf{q}_{s}&=&\mathbf{p}_{i}\mathbf{-p}_{f}+s\mathbf{k-p}_{B})=\frac{
16\pi }{(\mathbf{q}_{s}^{2}+1)^{2-i/p_{B}}}
\frac{\mathbf{\Delta }_{s}^{NR}.[%
\mathbf{\Delta
}_{s}^{NR}-\mathbf{p}_{B}.(1\mathbf{+}i\mathbf{/}p_{B})]}{[(
\mathbf{\Delta }_{s}^{NR})^{2}-(p_{B}+i)^{2}]^{1+i/p_{B}}}.
 \end{eqnarray}
In the expressions of the last two non relativistic TDCSs, the
argument of the ordinary Bessel functions is given by

\begin{equation}
z_{NR}=\frac{\mid a\mid }{cw}\mid \mathbf{\Delta }_{s}^{NR}\mid.
\end{equation}

\section{The TDCS in presence of the laser field. The DVPWBA2}

We now take into account the electronic relativistic dressing of
all electrons which are described by Dirac-Volkov plane waves
normalized to the volume $V$. This will give rise to a new trace
to be calculated but it will turn out that taking into account
the relativistic electronic dressing of the ejected electron
amounts simply to introduce a new sum on the $l_{B}$ photons that
can be exchanged with the laser field. The transition amplitude
in the DVPWBA2 is now given by

\begin{eqnarray}
&&S_{fi}=-\frac{i}{c}\int_{-\infty }^{\infty }dx^{0} <\psi
_{q_{f}}(x_{1})\phi _{f}(x_{2})\mid V_{d}\mid \psi
_{q_{i}}(x_{1})\phi _{i}(x_{2})>.
\end{eqnarray}
The difference between DVPWBA1 and DVPWBA2 is related to the way we choose $%
\phi _{f}(x_{2})$. Now, the Dirac-Volkov wave function for the
ejected electron is such that

\begin{equation}
\phi _{f}(x_{2})=\psi _{q_{B}}(x_{2})=[1+\frac{\ks\As_{(2)}}{2c(k.p_{B})}]\frac{%
u(p_{B},s_{B})}{\sqrt{2Q_{B}V}}e^{is_{B}(x_{2})},
\end{equation}
where $A_{(2)}=a_{1}\cos (\phi _{2})+a_{2}\sin (\phi _{2})$ is
the four potential of the laser field felt by the ejected
electron, $\phi
_{2}=k.x_{2}=k_{0}x_{2}^{0}-\mathbf{k.x}_{2}=wt-\mathbf{k.x}_{2}$
is the phase of the laser field and $w$ its pulsation. Proceeding
along the same line as before, we get for the unpolarized TDCS

\begin{eqnarray}
&&\frac{d\overline{\sigma }}{dE_{B}d\Omega_{B}d\Omega_{f}}=
\left.\sum_{s,l_{B}=-\infty }^{\infty}\frac{d\overline{\sigma
}^{(s,l_{B})}}{dE_{B}d\Omega _{B}d\Omega _{f}}\right|_{
Q_{f}=Q_{i}+(s+l_{B})w+\varepsilon _{b}-Q_{B}},
\end{eqnarray}

with

\begin{eqnarray}
\frac{d\overline{\sigma }^{(s,l_{B})}}{dE_{B}d\Omega _{B}d\Omega
_{f}} &=&
\frac{1}{2}\frac{|\mathbf{q}_{f}||\mathbf{q}_{B}|}{|\mathbf{q}_{i}|c^{6}}\frac{(\sum_{s_{i},s_{f}}\mid
M_{fi}^{(s)}\mid ^{2}/2)}{\mid
\mathbf{q}_{f}\mathbf{-q}_{i}\mathbf{-}s \mathbf{k}\mid
^{4}}\sum_{s_{B}}\mid \overline{u}(p_{B},s_{B})\Gamma
_{l_{B}}\gamma^{0}\mid ^{2}\nonumber\\ &\times&
\mid\Phi_{1,1/2,1/2}(\mathbf{q=\Delta
}_{s+l_{B}}\mathbf{-q}_{B})-\Phi
_{1,1/2,1/2}(\mathbf{q=-q}_{B}+l_{B}\mathbf{k})\mid ^{2}.
\end{eqnarray}

The quantity $\mathbf{\Delta }_{s+l_{B}}$ is simply given by $\mathbf{\Delta }
_{s+l_{B}}=\mathbf{q}_{i}\mathbf{-q}_{f}+(s+l_{B})\mathbf{k}$.
Introducing the factor $c(p_{B})=1/(2c(k.p_{B}))$, the symbol
$\Gamma _{l_{B}}$ is defined as
\begin{equation}
\Gamma _{l_{B}}=B_{l_{B}}(z_{B})+c(p_{B})[\as_{1}\ks
B_{1l_{B}}(z_{B})+\as_{2}\ks B_{2l_{B}}(z_{B})],
\end{equation}
where the three quantities $B_{l_{B}}(z_{B})$, $B_{1l_{B}}(z_{B})$ and $
B_{2l_{B}}(z_{B})$ are respectively given by

\begin{eqnarray}
\left\{\begin{array}{c}
B_{l_{B}}(z_{B})\\
B_{1l_{B}}(z_{B})\\
B_{2l_{B}}(z_{B})\end{array}\right\}=
\left\{\begin{array}{c}
J_{l_{B}}(z_{B})e^{il_{B}\phi _{0B}}\\
\{J_{l_{B}+1}(z_{B})e^{i(l_{B}+1)\phi
_{0B}}+J_{l_{B}-1}(z_{B})e^{i(l_{B}-1)\phi _{0B}}\}/2\\
\{J_{l_{B}+1}(z_{B})e^{i(l_{B}+1)\phi
_{0B}}-J_{l_{B}-1}(z_{B})e^{i(l_{B}-1)\phi
_{0B}}\}/2i\end{array}\right\}, \label{24}
\end{eqnarray}

where $z_{B}=\frac{|a|}{ c(k.p_{B})}\sqrt{(%
\widehat{\mathbf{y}}\mathbf{.p}_{B})^{2}+(\widehat{\mathbf{x}}\mathbf{.p}
_{B})^{2}}$
 is the argument of the ordinary Bessel functions
that will appear in the calculations and the phase $\phi _{0B}$
is defined by

\begin{equation}
\phi _{0B}=\arctan ((\widehat{\mathbf{y}}\mathbf{.p}_{B})/(\widehat{\mathbf{x%
}}\mathbf{.p}_{B})).
\end{equation}
The sum over the spins of the ejected electron can be transformed
to traces of gamma matrices. Using REDUCE \cite{21}, we find

\begin{eqnarray}
&&\sum_{s_{B}} \mid \overline{u}(p_{B},s_{B})\Gamma _{l_{B}}\gamma
^{0}\mid ^{2}=4\{E_{B}J_{l_{B}}^{2}(z_{B}) -wc(p_{B})(\cos (\phi
_{0B})(a_{1}.p_{B})+\sin (\phi
_{0B})(a_{2}.p_{B}))\nonumber\\&\times&
J_{l_{B}}(z_{B})(J_{l_{B}+1}(z_{B})+J_{l_{B}-1}(z_{B}))
-a^{2}w(k.p_{B})c^{2}(p_{B})(J_{l_{B}+1}^{2}(z_{B})+J_{l_{B}-1}^{2}(z_{B}))\}.
\end{eqnarray}
In  absence of the laser field, only the term
\\$4E_{B}J_{l_{B}}^{2}(z_{B}=0) \delta _{l_{B},0}=4E_{B}$
contributes, which was to be expected. Once again, one encounters
terms proportional to $\cos (\phi _{0B})$ as well as to $\sin
(\phi _{0B})$ which contributes to the sum over the spins of the
ejected electron. We compare this TDCS with the corresponding
cross sections in the framework of the NRPWBA2 (where the
incident, scattered and ejected electrons are described by non
relativistic Volkov plane waves )

\section{Results and discussion}

\subsection{In absence of the laser field}

We begin our discussion by the kinematics of the problem. In
absence of the
laser field, there is no dressing of angular coordinates \cite{22}  and we choose a geometry where $\mathbf{p}_{i}$ is along
the $Oz$
axis ($\theta _{i}=\phi _{i}=0$). For the scattered electron, we choose ($%
\theta _{f}=45^{\circ },$ $\phi _{f}=0$) and for the ejected
electron we
choose $\phi _{B}=180^{\circ }$ and the angle $\theta _{B}$ varies from $%
0^{\circ }$ to $360^{\circ }$. This is an angular situation where
we have a coplanar geometry. Spin effects are fully included and
we use an exact relativistic description of the electrons and the
atomic target. In order to check the validity of the coplanar
binary geometry, we begin first by comparing the three TDCSs
(RPWBA, NRPWBA and NRCBA) in the non relativistic
domain. In the expression of the Fourier transforms $\Phi _{1,1/2,1/2}(%
\mathbf{q=\Delta -p}_{B})$ and $\Phi
_{1,1/2,1/2}(\mathbf{q=-p}_{B})$, one
has to determine the angle between $\mathbf{\Delta -p}_{B}$ and $\mathbf{p}%
_{i}$ on the one hand and the angle between $\mathbf{-p}_{B}$ and $\mathbf{p}%
_{i}$ on the other hand. Keeping in mind that $\mathbf{p}_{i}$ is along the $%
Oz$ axis, one finds

\begin{equation}
\cos (\widehat{\mathbf{\Delta
-p}_{B},\mathbf{p}_{i}})=\frac{(p_{i}-p_{f}\cos \theta _{f}-p_{B}\cos
(\theta _{B}))}{\mid \mathbf{\Delta -p}_{B}\mid},
\end{equation}
while for $\cos \widehat{(-\mathbf{p}_{B},\mathbf{p}_{i})}=-\cos
(\theta _{B})$ from which one deduces the corresponding angles.
To have an idea of how the ejected electron ''looses'' its
Coulombian behaviour, we begin by a process where the incident
electron kinetic energy is $T_{i}=1350$ $eV$, and the ejected
electron kinetic energy is $T_{B}=574.5$ $eV$.

\begin{figure}[ht]
\centering
\includegraphics[angle=0,width=8.5cm,height=6.5cm]{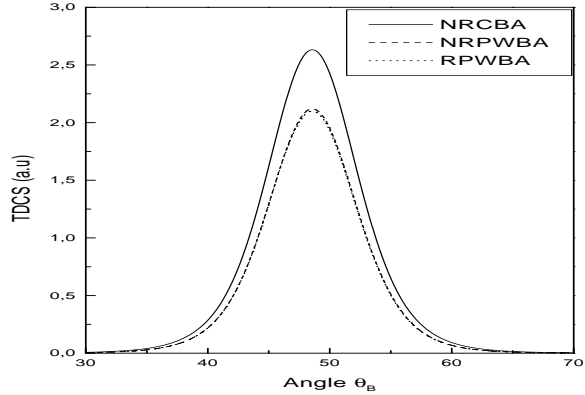}
\caption{The three TDCSs scaled in units of $10^{-3}$ $a.u$. The solid line represents the non relativistic TDCS in the Coulomb Born Approximation the long dashed line represents the corresponding TDCS in the plane wave approximation and the dotted line sketches the relativistic TDCS in the plane wave approximation. The incident electron kinetic energy is $T_i=1350\quad ev$ and the ejected electron kinetic energy is $T_B=574.5\quad ev$   }.
\end{figure}

 In Fig 1, we see
that both RPWBA and NRPWBA give nearly the same results whereas
NRCBA gives a higher TDCS due to the fact that the ejected
electron still feels the influence of the atomic field.
Increasing this energy of the ejected electron from $574.5$ $eV$
to $674.5$ $eV$ gives rise to almost three indistinguishable
curves. This situation is shown in Fig 2.
\begin{figure}[ht]
\centering
\includegraphics[angle=0,width=8.5cm,height=6.5cm]{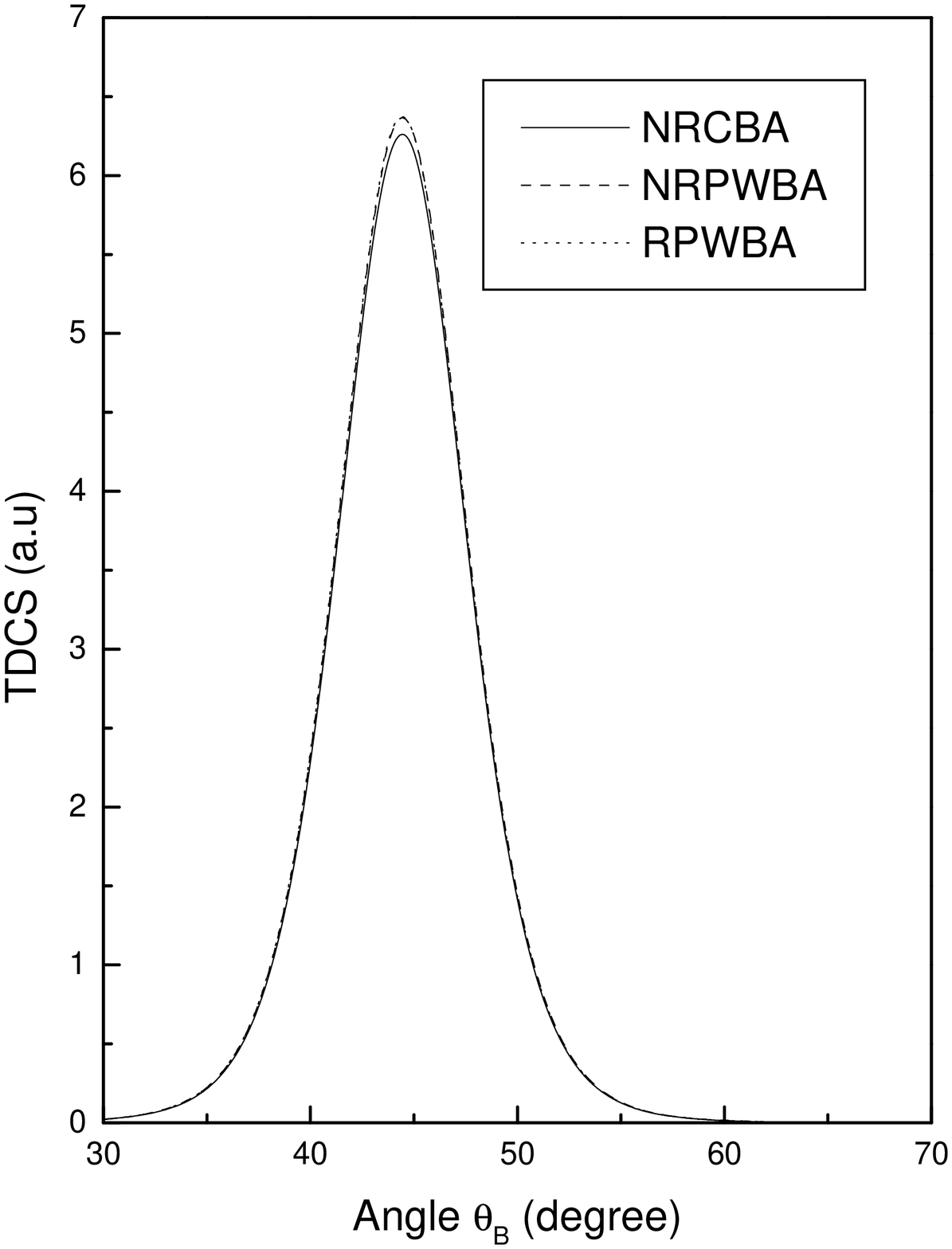}
\caption{Same as in Fig. 1 but for $T_i=1350\quad ev$ $T_B=674.5\quad ev$  }.
\end{figure}As the incident
electron kinetic energy is
increased to $2700$ $eV$ which corresponds to a relativistic parameter $%
\gamma =(1-v^{2}/c^{2})^{-1/2}=1.0053$, that of the ejected electron to $%
1349.5$ $eV$, we have a very good agreement between the three
TDCSs. This is a crucial test for the models we will develop in
the presence of a laser field since the adjunction of the latter
cannot be done without a judicious choice of a geometry. It must
be borne in mind that this coplanar binary geometry is not well
suited for the study of the domain of low kinetic energy for the
ejected electron. The agreement between these three approaches
remains good up to $T_{i}=15$ $keV$ from which the RPWBA gives
results that are a little lower than NRPWBA and NRCBA because
relativistic and spin effects can no longer be ignored. Due to
the relative simplicity of the model (even if the calculations
are far from being obvious) and the exact non dressed
relativistic description of the target, it is remarkable that
such an agreement should be reached for these energies. When the
laser field is introduced, the dressing of angular coordinates is
not important for the non relativistic regime ($\gamma =1.0053$,
$\mathcal{E}=0.05$ $a.u)$
but becomes noticeable for the relativistic regime ( $\gamma =2.0$, $%
\mathcal{E}=1.00$ $a.u$) where $\mathcal{E}$ is the electric
field strength. The unit of electric field strength in atomic
units is $\mathcal{E=}5.14225$ $10^{9}$ $V/cm$.
\subsection{In presence of the laser field}

\subsubsection{A. The non relativistic regime ($\gamma =1.0053$, $\mathcal{E}%
=0.05$ $a.u)$}

The first check to be done is to take a zero electric field
strength in order to recover all the results in absence of the
laser field. We have done these checks for all approximations. It
has been shown \cite{23} that for a laser
frequency $w=0.043$ $a.u$, which corresponds to a laser photon energy of $%
1.17$ $eV$, dressing effects due to the atomic target are not very
important. A complete and exact relativistic treatment of the
ejected electron is not analytically possible since the non
relativistic wave equation for continuum states in a Coulomb
field is separable in parabolic coordinates, but the
corresponding Dirac equation is not. In other words, a
decomposition of the relativistic continuum wave function into
partial waves is not as straightforward as for the non
relativistic case and the quantum numbers of each partial wave
have to be taken into account very carefully. However, a tedious
numerical construction of the first few partial waves is possible.

We first compare the results obtained within the three
approximations (DVPWBA1, NRPWBA1, NRCBA1) where it is expected on
physical grounds that these cannot be used to study the
relativistic regime. The three summed
TDCSs are all peaked around $\theta _{B}=45{{}^{\circ }}$, $\phi _{B}=180{%
{}^{\circ }}$ which was to be expected for the case of the
geometry chosen
since for the scattered electron, the choice we have made is $\theta _{f}=45{%
{}^{\circ }}$, $\phi _{f}=0{{}^{\circ }}$ and in the $xOy$ plane,
this amounts to a scattered electron and an ejected electron
having an opposite value of $\theta $. Even with no photon
exchange and for an electric field stregth of $0.05$ $a.u$, the
presence of the laser field reduces considerably the magnitude of
the TDCSs. The NRPWBA1 and NRCBA1 TDCSs are nearly
indistinguishable whereas the DVPWBA1 TDCS is lower than the
former ones in the vicinity of the maximum for $\theta
_{B}=45{{}^{\circ }}$. For this angle, we have TDCS(NRPWBA1)
$\simeq 0.717$ $10^{-5}$ $a.u$,
TDCS(NRCBA1) $\simeq 0.7115$ $10^{-5}$ $a.u$, TDCS(DVPWBA1) $\simeq 0.472$ $%
10^{-5}$ $a.u$. Two interesting cases are those corresponding to
the absorption and emission of one photon. We have shown in a
previous work \cite{24}  that when
all electrons are described by Dirac-Volkov planes, the
corresponding differential cross sections for the absorption and
emission (of one photon) processes are identical. It is not the
case for these three approximations since the ejected electron is
described by a free Dirac plane wave. The DVPWBA1 TDCS is larger
than the two other non relativistic TDCSs by a factor $4$ at the
maximum for $\theta _{B}=45{{}^{\circ }}$ for the absorption
process and the emission process but these relativistic TDCSs are
not identical . The TDCS for the emission of one photon is
smaller than the corresponding one for the absorption of one
photon by a factor $2$ at the same maximum. These remarks are not
without interest since a crucial test of our next model ( all
electrons are described by Dirac-Volkov plane waves) will be to
compare the two TDCSs within the framework of DVPWBA2 for these
two processes. It will be a sound consistency check of our
calculations.
\begin{figure}[ht]
\centering
\includegraphics[angle=0,width=8.5cm,height=6.5cm]{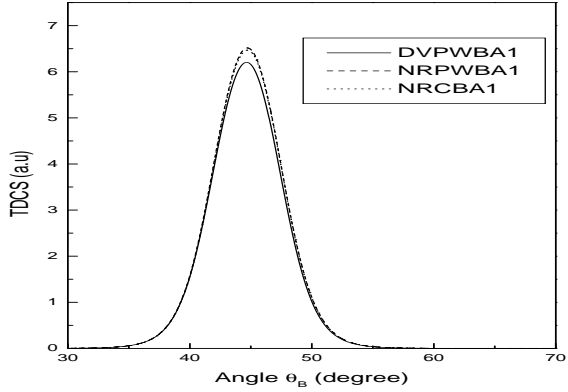}
\caption{The summed TDCSs for an exchange of $\pm 100$ photons in non relativistic regime scaled in units of $10^{-4} $ $a.u$}.
\end{figure}
In Fig. 3, we show the three summed TDCSs for an
exchange of $\pm 100$ photons and we obtain close curves for the
non relativistic plane wave and Coulomb Born results with the
relativistic plane wave TDCS a little lower than the former ones
in the vicinity of the peak at $\theta _{B}=45{{}^{\circ }}$. For
angles lower or larger than $\theta _{B}=45{{}^{\circ }}$, the
three TDCSs give nearly the same results. The dressing of angular
coordinates being negligeable, the maximum is maintained for the
above mentionned value of $\theta _{B}$ but in the relativistic
regime, this maximum is shifted. Also, we have compared the
relativistic TDCS for different numbers of photons exchanged,
typically $\pm 50$, $\pm 100$ and $\pm 150$ photons. The TDCSs
increase when the number of the photons exchanged increases and
finally, we have compared the relativistic TDCS without laser
field with these summed relativistic TDCSs in order to obtain a
check of the well known pseudo sumrule \cite{25}. We have obtained
results that converge to the relativistic TDCS without laser field
but complete convergence is not reached since the ejected
electron is not properly described.

Now, we discuss the results obtained within the framework of the
three more accurate approximations (DVPWBA2, NRPWBA2, NRCBA2) in
the same non relativistic regime. The description of the ejected
electron by a Dirac-Volkov plane wave being more accurate and
necessary on physical grounds (there are no constraints that
forbid the ejected electron to exchange photons with the laser
field), we have first investigated the case where no photon is
exchanged at all ($s=0,l_{B}=0$). There is a shift of the
location for the maximum corresponding to the relativistic TDCS
while the two non relativistic TDCSs are nearly the same. The
magnitude of the TDCSs is also considerably reduced. We have
TDCS(DVRPWBA2) $\simeq 0.347$ $10^{-8}$ $a.u$ for $\theta
_{B}=41{{}^{\circ }}$ while TDCS(NRPWBA2) $\simeq $
TDCS(NRCBA2) $\simeq 0.3326$ $10^{-8}$ $a.u$ for $\theta _{B}=43{{}^{\circ }}$%
. There are three small secondary peaks for the relativistic TDCS
and two secondary peaks for the two non relativistic TDCSs.This
behaviour stems in the relativistic description from the
contribution of the sum over the spins
of the ejected electron. This sum shows a narrow peak for $\theta _{B}=41{%
{}^{\circ }}$ and presents two minima for $\theta _{B}=45{{}^{\circ }}$ and $%
\theta _{B}=50{{}^{\circ }}$. The other peaks can also be traced
back to the behaviour of this sum. Relativistic and spin effects
begin at this stage to become noticeable since the shift as well
as the magnitude of the relativistic TDCS with respect to the non
relativistic TDCSs are clear signatures even in the non
relativistic regime.

Moreover, the most crucial test of our model is the complete
symmetry between the emission and absorption
processes.
\begin{figure}[ht]
\centering
\includegraphics[angle=0,width=8cm,height=6.5cm]{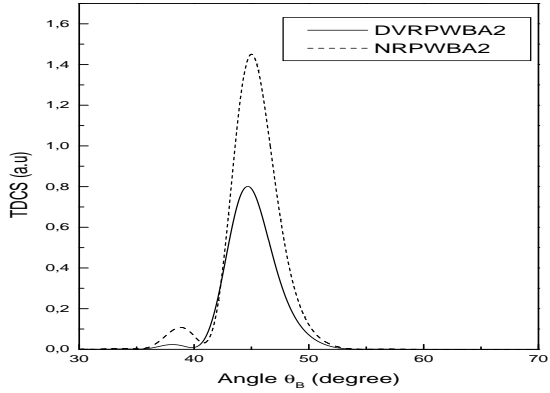}
\caption{The two TDCSs for $s=1$ and $l_B=-1$ in the non relativistic regime scaled in units of $10^{-8}$ $a.u$. We obtain the same figure for  $s=-1$ and $l_B=1$}
\end{figure}
In Fig. 4,
we show the relativistic and non relativistic TDCSs for $s=1$ and
$l_{B}=-1$. We have obtained the same curve for the case $s=-1$
and $l_{B}=1$. For these two curves, the maximum shifts back to
the value $\theta _{B}=45{{}^{\circ }}$. As the number of
exchanged photons $s$ is linked both to the incident and
scattered electron, we simulated a process where the number
$s=2l_{B}$ to check the corresponding influence on the TDCSs and
on the location of the maximum. The summed TDCSs for ($s=\pm 50$,
$l_{B}=\pm 25$) and for ($s=\pm 100$, $l_{B}=\pm 50$) are almost
halved when compared to the corresponding TDCSs for $s=l_{B}$.
All curves are peaked around a maximum angle due to the behaviour
of the square of the Fourier transform of the relativistic
hydrogenic wave functions that falls off rapidly to zero in a
small region around this maximum. Another interesting remark
concerns once again the behaviour of the sum over the spins of
the ejected electron with the number of photons exchanged. As
$l\geq \pm 500$, this sum is almost zero and contributes also to
the rapid fall off of the corresponding relativistic TDCS. To
illustrate the complexity of the location of the visual cut-off of
the relativistic TDCS, we consider an electric field strength $\mathcal{E}
=0.01$ $a.u$ where it is expected that the numbers $s$ and $l$ of
photons exchanged will not be high. The envelope of photon energy
transfer obtained is a curve in three dimensions. As in the case
of the process of Mott scattering in a strong laser field, we
observe in Fig. 5

\begin{figure}[ht]
\centering
\includegraphics[angle=0,width=8cm,height=6.5cm]{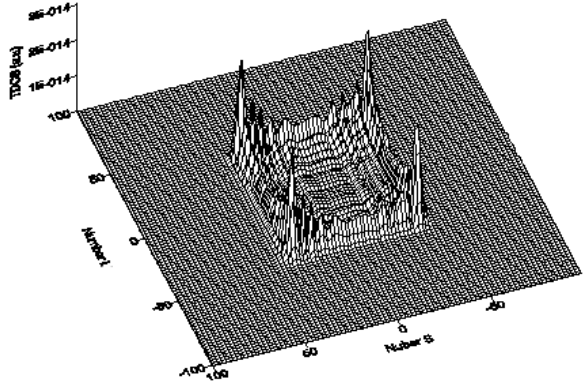}
\caption{The enveloppe of photon energy transfer in the non relativistic regime and for an electric field strength $\mathcal{E}=0.01\,a.u$}
\end{figure}

a rapid fall off of the relativistic TDCS for
$s\simeq l\simeq \pm 40$ where the absolute value of the indices
of the ordinary Bessel functions are close to their arguments.
Also, as a side result, we see clearly in this Figure that there
is a complete symmetry between $s$ and $l$. To obtain a
converging envelope, one has to sum over the same numbers $s$ and
$l$ of photons exchanged.

\subsubsection{B. The relativistic regime ($\gamma =2.$, $\mathcal{E}=1.00$ $%
a.u)$}

In the relativistic regime, dressing of angular coordinates is not
negligeable and we indeed observed as in the case of excitation a
shift from
$\theta _{B}=45{{}^\circ}$ to lower values. The relativistic parameter $%
\gamma =2$ corresponds to an incident electron kinetic energy of
$c^{2}$ in atomic units or the rest mass of the electron ($0.511$
$MeV$).
\begin{figure}[ht]
\centering
\includegraphics[angle=0,width=8cm,height=6.5cm]{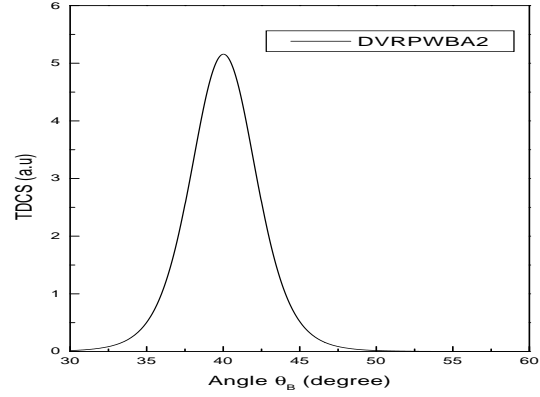}
\caption{The RPWBA  TDCS without laser field in the relativistic regime,  scaled in $10^{-16}\quad a.u$}
\end{figure}
 In Fig 6, we show the behaviour of the relativistic TDCS
in absence of a laser field in order to have an idea about the
reduction of the corresponding TDCS when the
laser field is introduced. The maximum is well located at $\theta _{B}=40{
{}^\circ}$ and at this value, we have for the corresponding TDCS(RPWBA)
$\simeq 0.54$ $10^{-15}$ $a.u$. For this regime and in presence of a
strong laser field, the non relativistic TDCSs are no longer
reliable and we will focus instead on the discussion of the
results obtained within the DVPWBA1 and DVPWBA2.
For the DVPWBA1, we first analyzed what happens when no photon is
exchanged. There is a drastic reduction of the TDCS with a
maximum shifted for $\theta _{B}$ lower than $40{{}^\circ}$ and
at this value, we have for the corresponding TDCS(DVPWBA1)
 $\simeq 0.5$ $10^{-21}$ $a.u$.
Once again, there is an asymmetry
between the absorption and emission processes of one photon. Due
to a lack of high speed computing facilities we cannot check the
pseudo sumrule but we have a TDCS that increases when the number
of the photons exchanged increases.
For the DVPWBA2, there is also a considerable reduction of the TDCS for $
s=l_{B}=0$.
\begin{figure}[h]
\centering
\includegraphics[angle=0,width=8cm,height=6.5cm]{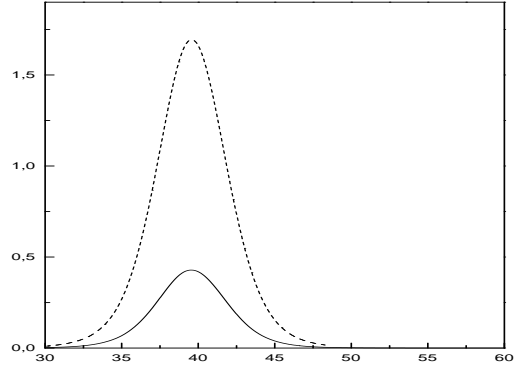}
\caption{The summed TDCS (a) for an exchange of $s=\pm 50$, $l_{B}=\pm 50$ (b) for an exchange of $s=\pm 100$, $l_{B}=\pm 100$ in the relativistic regime,  scaled in $10^{-22}\quad a.u$ }
\end{figure}
 In Fig 7, we compare the summed relativistic TDCS for ($s=\pm 50$%
, $l_{B}=\pm 50$) and for ($s=\pm 100$, $l_{B}=\pm 100$) where
the shift of the maximum is clearly visible. We still have a
complete symmetry between the absorption and emission processes.

\subsection{Conclusion}

In this work, we have investigated the contribution of the
relativistic electronic dressing in laser-assisted ionization of
atomic hydrogen by electronic impact using the Dirac-Volkov plane
wave solutions to describe the incoming and the two outgoing
electrons. We have worked in the binary coplanar geometry where
the description of the ejected electron by a relativistic Coulomb
wave function is not necessary. The influence of the laser field
is taken into account to all orders in the Dirac-Volkov
description of electrons and the description of the atomic target
used is the analytical relativistic hydrogenic wave functions. It
turns out that all the TDCSs are well peaked around a maximum
angle due to the behaviour of the Fourier transforms of the
relativistic hydrogenic wave functions. Symmetry between the
absorption and emission processes is obtained when all electrons
are described by Dirac-Volkov plane waves both for the non
relativistic and relativistic regime.

\end{document}